\documentstyle[12pt, epsfig]{article}            %%%
\newfont{\ffont}{msym10}                          %%
\newcommand{\beq}{\begin{equation}}               %%
\newcommand{\eeq}{\end{equation}}                 %%
\newcommand{\bqry}{\begin{eqnarray}}              %%
\newcommand{\eqry}{\end{eqnarray}}                %%
\newcommand{\bqryn}{\begin{eqnarray*}}            %%
\newcommand{\eqryn}{\end{eqnarray*}}              %%
\newcommand{\preprint}[1]{\begin{table}[t]        %%
            \begin{flushright}                    %%
            \begin{large}{#1}\end{large}          %%
            \end{flushright}                      %%
            \end{table}}                          %%
\newcommand{\PD}[2]                               %%
    {\frac{\partial^{#2}}{\partial #1^{#2}}}      %%
\newcommand{\eproof}{\hspace{0.3cm} \mbox{\boldmath  $\Box $ } }

\newcommand{\btab}{\begin{tabbing}}
\newcommand{\etab}{\end{tabbing}}

\newcommand{\beqn}{\begin{equation}}
\newcommand{\eeqn}{\end{equation}}
\newcommand{\barr}[1]{\begin{array}{#1}}
\newcommand{\earr}{\end{array}}
\newcommand{\beqna}{\begin{eqnarray}}
\newcommand{\eeqna}{\end{eqnarray}}
\newcommand{\btablec}{\begin{table} \begin{center}}
\newcommand{\etablec}{\end{center} \end{table}}

\newcommand{\gapproxeq}{\lower.7ex\hbox{$\;\stackrel{\textstyle>}
{\sim}\;$}}
\newcommand{\plabel}[1]{\label{#1}}
\newcommand{\pbibitem}[1]{\bibitem{#1}}
\catcode`\@=11 \@addtoreset{equation}{section}    %%
\renewcommand{\theequation}                       %%
         {\arabic{section}.\arabic{equation}}     %%
\begin{document}
\preprint{\small{hep-ph/9807400} \\ \small{LA-UR-98-3047}}
\title{Scalar Glueball Mixing and Decay}
\author{\\ Leonid Burakovsky\thanks{E-mail: BURAKOV@T5.LANL.GOV} \ and \
Philip R. Page\thanks{E-mail: PRP@T5.LANL.GOV}
 \\  \\  Theoretical Division, MS B283 \\  Los Alamos National Laboratory \\ 
Los Alamos, NM 87545, USA \\}
\date{ }
\maketitle
\begin{abstract}
We provide the first explanation of the counter--intuitive scalar glueball 
couplings to pseudoscalar mesons found in lattice QCD and predict hitherto 
uncalculated decay modes. Significant $a_1\pi $ and $(\pi \pi )_S(\pi \pi )_S$
couplings are found. We demonstrate the equivalence of linear and quadratic 
mass matrices for glueball--quarkonium mixing. The equivalence of formalisms 
which deal with a glueball--quarkonium basis and only a 
quarkonium basis is demonstrated. We show that the $f_0(1500)$ is not the 
heaviest state arising from glueball--quarkonium mixing for a glueball mass 
consistent with lattice QCD. The masses and couplings of scalar mesons, as 
well as their valence content, are calculated.
\end{abstract}

\bigskip
\section{Introduction}
The existence of a gluon self--coupling in QCD suggests that, in addition to 
the conventional $q\bar{q}$ states, there may be non--$q\bar{q}$ mesons: bound 
states built from gluons, called glueballs. 
%However, the theoretical quidance 
%on the properties of unusual states is often contradictory, and models that 
%agree in the $q\bar{q}$ sector differ in their predictions about new states. 
The abundance of $q\bar{q}$ meson states in the 1--2 GeV region and 
the possibility of glueball--quarkonium mixing makes the identification of the 
would--be lightest non--$q\bar{q}$ mesons extremely difficult. To date, no 
glueball state has been firmly established, although the existence of 
glueballs has been established in lattice QCD.

Although the current situation with the identification of glueball states is 
rather complicated, some progress has been made recently in the
scalar glueball sector, where both experimental and lattice QCD 
results seem to converge. Various lattice QCD glueball mass estimates have 
been made in the literature, and one of the differences stems from the way the
physical results are obtained from the raw lattice data: either by calculating
the sting tension, or the $\rho$ mass. UKQCD estimates $1.55\pm 0.05$ GeV by 
fixing to the string tension \cite{bali93}. GF11 originally estimated 
$1.74\pm 0.07$ GeV \cite{weinmass} by fixing to the $\rho$ mass. Later 
estimates include $1.71\pm 0.06$ GeV \cite{weinmix} and $1.65\pm 0.06$ GeV
\cite{LW1}. Attempts at reconciling UKQCD and GF11 results yielded 
$1.57\pm 0.09$ \cite{teper96} and $1.63\pm 0.09$ \cite{weinmix}. 
%Also, a 
%na\"{\i}ve estimate for the scalar glueball mass in ref. \cite{glue}, based on
%Regge phenomenology, leads to $1.62\pm 0.01$ GeV. 
In what follows, we shall 
take the glueball mass $m_G$ to be 1.6 GeV. 
Accordingly, there are two experimental candidates \cite{pdg}, $f_0(1500)$ and
$f_J(1710),$ in the right mass range.
%, and this may be considered as strong evidence
%for one of these states being a scalar glueball.

Recently, ref. \cite{BBG} showed that the hypothesis where isoscalar meson
mixing proceeds through an intermediate glueball, called ``glueball 
dominance'', can consistently account for isoscalar meson masses in various
$J^{PC}$ sectors by employing glueball masses predicted by lattice QCD. Here 
we explore in detail the consequences of glueball dominance in the scalar 
sector, which differs from any other $J^{PC}$ sector due to the relative 
proximity of the glueball and quarkonia masses. Particularly, we demonstrate 
that the formulation of glueball dominance in refs. \cite{BBG,HK} is consistent
with glueball--quarkonium mixing formulated in refs. 
\cite{weinmix,LW1,LW2,Wein}. Moreover, the recent lattice QCD 
calculation of ref. 
\cite{weindecay} found an unusual decay pattern for the scalar glueball, which
is not consistent with any published model. We 
demonstrate that glueball dominance can explain this decay pattern. 

In Section 2 the canonical formulation of glueball dominance is introduced.
Section 3 merges glueball dominance and the $^3P_0 /$ flux--tube 
model to describe glueball 
decays. Section 4 discusses glueball--quarkonium mixing. Phenomenological 
implications are indicated in Section 5, and a summary given in Section 6.

\section{Glueball dominance}

We assume the glueball dominance of quarkonium mixing, viz., there is no 
direct quarkonium--quarkonium mixing, and the $q\bar{q}\leftrightarrow 
q^{'}\bar{q}^{'}$ transition is dominated by the glueball with the 
corresponding quantum numbers in the intermediate state \cite{Hou}.

Although the validity of glueball dominance has not been shown in QCD, 
the conclusions drawn from glueball dominance often agree with
conclusions derived from the large number of colors $N_c$ limit of QCD:

\begin{itemize}

\item  Consider a Feynman graph where a $q\bar{q}$ pair annihilates into
an arbitrary number of gluons which then create a $q\bar{q}$ pair. In the
large $N_c$ limit the graph is ${\cal O}(\frac{1}{N_c})$ independent of the
number of intermediate gluons. This corresponds to the finding (in glueball 
dominance) that the coupling between mesons via an intermediate glueball
is largely independent 
of the C--parity of the glueball \cite{BBG}, i.e. the number of gluons that a glueball can be 
built from in perturbative QCD. 

\item Consider a Feynman graph where (a) two gluons each create
a $q\bar{q}$ pair (i.e. two quark loops) which again combines into two gluons;
(b) two gluons combine to a single intermediate quark loop and then combine to
two gluons. (a) is ${\cal O}(\frac{1}{N_c^2})$ and (b) is ${\cal O}(\frac{1}{
N_c})$. Glueball dominance postulates that glueballs mix via a single 
intermediate meson, corresponding to (b), in agreement with the large $N_c$ 
limit.

\item The glueball dominance description of glueball decay postulates
that the glueball couples to a meson, which then subsequently decays. In order
for this process to happen, the time $\tau$ for the $q\bar{q}$ pair created in 
the glueball to form a meson should be significantly less than the time
${1}/{\Gamma}$ required for yet another $q\bar{q}$ pair to form so that the
meson decays. In the large  $N_c$ limit, ${1}/{\Gamma}$ is ${\cal O}(N_c)$.
The time taken for the created $q\bar{q}$ pair in the glueball to form a meson
should be inversely proportional to the mass of the state, so that
$\tau$ is ${\cal O}(1)$. Hence the large  $N_c$ limit gives the condition
$\tau \ll {1}/{\Gamma}$ required by the glueball dominance. Another way to 
see the large $N_c$ result is to note that quark pair creation is 
suppressed (${\cal O}(\frac{1}{N_c})$) but quark rearrangement is not
(${\cal O}(1)$), so that created quarks in a glueball would rearrange
to form a meson, rather than create another $q\bar{q}$ pair.

\end{itemize}

Here we review the glueball dominance picture discussed in more detail in 
ref. \cite{BBG}. The possibility of the transition $q\bar{q}\rightarrow gg
\ldots g\rightarrow q^{'}\bar{q}^{'}$
% with two gluons in the color singlet state for the 
%pseudoscalar mesons, three gluons for the vector mesons, etc., 
is accounted for by the quark mixing amplitudes, $A_{qq^{'}},$ which are 
included in the meson mass square matrix (written down here in the  
$s\bar{s}$,  $(u\bar{u}+d\bar{d})/\sqrt{2}$ basis):
\beq
\left( 
\begin{array}{cc}
m^2_{s\bar{s}}+A_{ss} & \sqrt{2}A_{sn} \\
\sqrt{2}A_{ns} & m^2_{n\bar{n}}+2A_{nn}
\end{array}
\right) ,
\eeq
where $m_{s\bar{s}}$ and $m_{n\bar{n}},\;n=u,d$  are the primitive (bare)
quarkonia masses.

The quark mixing amplitudes can be represented in the standard form,
$$A\sim ({\rm Vertex}1)\cdot ({\rm Propagator})\cdot ({\rm Vertex2}),$$ 
which reduces to
\beq
A_{qq^{'}}=\sum _i\frac{\langle q\bar{q}|H_{p.c.}^{q^{'}\bar{q}^{'}}|i\rangle 
\langle i|H_{p.c.}^{q\bar{q}}|q^{'}\bar{q}^{'}\rangle }{M^2(q^{'}\bar{q}^{'})-
M^2(i)},
\eeq
where $H_{p.c.}^{q\bar{q}}$ is the quark pair creation operator for the flavor
$q,$ and $|i\rangle $ is a complete set of the (gluon) intermediate states. 
Because of the assumed glueball dominance, the sum (2.2) is saturated by the 
low--lying glueball:
\beq
A_{qq}\simeq \frac{f^2_{q\bar{q}G}}{m_{q\bar{q}}^2-m_{\tilde{G}}^2},\;\;\;
f_{q\bar{q}G}\equiv \langle q\bar{q}|H_{p.c.}^{q\bar{q}}|G\rangle \Big| _{
p^\mu p_\mu =m_{q\bar{q}}^2},
\eeq
for $q=n(=u,d),s,$ and, in view of the factorization hypothesis discussed in
more detail in ref. \cite{BBG},
\beq
A_{sn}=A_{ns}\equiv\sqrt{A_{nn}\cdot A_{ss}}\simeq \frac{f_{n\bar{n}G}f_{s\bar{
s}G}}{\sqrt{(m_{n\bar{n}}^2-m_{\tilde{G}}^2)(m_{s\bar{s}}^2-m_{\tilde{G}}^2)}},
\eeq
where $f_{q\bar{q}G}$ defined in (2.3) is the coupling of the intermediate 
glueball to $q\bar{q}$, and $m_{\tilde{G}}$ is the corresponding (physical) 
glueball mass. 

%In the $SU(3)$ limit, 
%\beq
%f_{n\bar{n}G}\equiv \langle u\bar{u}|H_{p.c.}^{u\bar{u}}|G\rangle =\langle d
%\bar{d}|H_{p.c.}^{d\bar{d}}|G\rangle =
%\langle s\bar{s}|H_{p.c.}^{s\bar{s}}|G\rangle \equiv f_{s\bar{s}G}.
%\eeq
%We assume this limit for simplicity, and therefore consider the following 
%glueball dominance mass matrix ansatz: 

Thus, the mass matrix (2.1) reduces to
\beq
\left(
\begin{array}{cc}
m_{s\bar{s}}^2+\frac{f_{s\bar{s}G}^2}{m_{s\bar{s}}^2-m^2_{\tilde{G}}} & 
\frac{\sqrt{2}
f_{n\bar{n}G}f_{s\bar{s}G}}{\sqrt{(m_{n\bar{n}}^2-m^2_{\tilde{G}})(
m_{s\bar{s}}^2-m_{\tilde{G}}^2)}} \\
\frac{\sqrt{2}f_{n\bar{n}G}f_{s\bar{s}G}}{\sqrt{(m_{n\bar{n}}^2-m^2_{
\tilde{G}})(m^2_{s\bar{s}}-m^2_{\tilde{G}})}} & 
m_{n\bar{n}}^2+\frac{2f_{n\bar{n}G}^2}{m_{n\bar{n}}^2-m_{\tilde{G}}^2}
\end{array}
\right) .
\eeq
%The mass matrix (2.5) was used as the basis of the analysis in ref. 
%\cite{BBG}. 
The masses of the physical isoscalar states $f_0$ and $f_0^{'}$ 
are obtained by diagonalizing this mass matrix:
\beq
\left(
\begin{array}{cc}
m^2_{f_0^{'}} & 0 \\
0 & m^2_{f_0}
\end{array}
\right) .
\eeq
%and, with $f^2\ll (m_G^2-m_{q\bar{q}}^2)^2,$ are given by the following 
%expressions ($m_{q\bar{q}}\equiv m_{n\bar{n}}\sim m_{s\bar{s}}):$
%\bqry
%m_{f_0^{'}}^2 & = & m_{s\bar{s}}^2\;-\;\frac{f^2}{m_G^2-m_{s\bar{s}}^2}\left[
%1\;+\;O\left( \frac{f^2}{(m_G^2-m_{q\bar{q}}^2)^2}\right) \right] , \\
%m_{f_0}^2 & = & m_{n\bar{n}}^2\;-\;\frac{2f^2}{m_G^2-m_{n\bar{n}}^2}\left[
%1\;+\;O\left( \frac{f^2}{(m_G^2-m_{q\bar{q}}^2)^2}\right) \right] .
%\eqry

\section{Glueball decay}
\subsection{Scalar glueball decay to two pseudoscalar mesons}
Glueball decay via glueball dominance was introduced in ref. \cite{hou84}. We 
follow their approach, except for one improvement. Ref. \cite{hou84} coupled 
the decaying glueball to an off--shell meson, which then subsequently decays to
the outgoing mesons. The coupling used for this latter process is extracted
from experiment where the off--shell meson is on--shell. This should not 
introduce undue errors unless the coupling is strongly dependent on energy of 
the off--shell meson.

In this section we shall deal with a case where the coupling is strongly 
dependent on the energy of the off--shell meson. Our proposed solution is to 
calculate the decay for the correct energy of the off--shell meson by employing
the phenomenologically successful $^3 P_0$ model \cite{kokoski87,geiger94}. 
Since the nonrelativistic $^3 P_0$ and Isgur--Paton flux--tube decay model 
give identical predictions for simple harmonic oscillator meson wave functions
\cite{page95charm}, which we employ, our predictions can also be viewed as 
predictions of the flux--tube model.

In fact, the flux--tube model, motivated from the strong coupling limit of the
Hamiltonian formulation of the lattice gauge theory (HLGT), added to glueball 
dominance affords an intuitive picture of the decay process. In the lowest 
order in perturbation theory glueballs can be viewed as rings of flux in HLGT.
Pair creation occurs in the first order of perturbation theory and breaks the 
flux--ring up into a meson. The flux--tube connecting the two quarks in the 
meson then breaks via the creation of a quark--antiquark pair with vacuum ($^3 
P_0$) quantum numbers to form two outgoing mesons \cite{paton85}. 

The amplitude (in GeV) for the decay of the scalar glueball to two outgoing 
mesons is then given by
\beqn
{\cal M} = \frac{\sqrt{2} f_{n\bar{n}G}}{m_{n\bar{n}}^2 - m_{\tilde{G}}^2} \; 
{\cal A}(n\bar{n}) + \frac{f_{s\bar{s}G}}{m_{s\bar{s}}^2 - m_{\tilde{G}}^2} \;
{\cal A}(s\bar{s}),
\eeqn
where $f_{n\bar{n}G}$ and $f_{s\bar{s}G}$ are the couplings of the
scalar glueball to the intermediate $n\bar{n}$ and $s\bar{s}$ scalar mesons,
respectively, introduced in Section 2. We have also taken care to insert the 
scalar meson propagator with the masses ordered according to the prescription
of glueball dominance (2.5). ${\cal A}(n\bar{n})$ and ${\cal A}(s\bar{s})$ are
the intermediate scalar meson $^3P_0$ model decay amplitudes
to two outgoing mesons. We have assumed that only the ground
state scalar mesons saturate the decay.

The full $^3P_0$ model amplitude is given in Appendix B of ref. 
\cite{kokoski87}; here we just write down the case of identical inverse 
radii $\beta_A = \beta_B = \beta_C \equiv \beta$ for identical 
quark masses\footnote{The $^3P_0$ model amplitudes depend 
explicitly on the light and strange quark masses for decays where the initial 
quarks are different from the quarks in the created pair. In this work 
we take the light and strange quark masses to be 
identical for decays to $K\bar{K}$.}, 
for simplicity (A denotes the scalar meson, and B and C the outgoing mesons):
\beqn \plabel{coupling}
{\cal A}  = \sqrt{\frac{8m_G^2\tilde{M}_B\tilde{M}_C}{\tilde{M}_A}}\;
\frac{16\pi^{\frac{3}{4}}}{9\sqrt{\beta}}\;\gamma_0\; (1 - 
\frac{2p^2}{9\beta^2})\;\exp \;\!\{-\frac{p^2}{12\beta^2}\}, 
\eeqn
where we neglected the factor arising from the flavors of the mesons. The mock 
meson phase space convention is specified by $\tilde{M}_{A,B,C}$ 
\cite{kokoski87}, and $p$ is the momentum of the outgoing meson B in the 
glueball rest frame. The width is computed from the amplitude in Eq. 
(\ref{coupling}) by using the standard formula given by the Particle Data 
Group \cite{pdg}. The pair creation constant $\gamma_0$ is usually taken to 
be the same for $n\bar{n}$ and $s\bar{s}$ pair creation 
\cite{kokoski87,biceps}, as we shall do here.

The composition of $\eta$ and $\eta^{'}:$ is
\beqn
\eta = \sin\theta\; |n\bar{n}\rangle + \cos\theta\; |s\bar{s}\rangle ,\;\;\; 
%\hspace{1cm}
\eta^{'} = \cos\theta\; | n\bar{n} \rangle - \sin\theta\; |s\bar{s}\rangle .
\eeqn

We obtain the following simple relationships between the $^3P_0$ model
amplitudes when  $\beta_A, \beta_B, \beta_C, p$ and the mock meson phase space
are taken to be constant
for all meson decay processes:
\beqna
\lefteqn{ {\cal A}(\frac{u\bar{u}+d\bar{d}}{\sqrt{2}}\rightarrow \pi \pi ) \; 
: \;  
{\cal A}(\frac{u\bar{u}+d\bar{d}}{\sqrt{2}}\rightarrow K\bar{K}) \; : \;  
{\cal A}(\frac{u\bar{u}+d\bar{d}}{\sqrt{2}}\rightarrow \eta \eta ) \; : \; 
\nonumber } \\ &  &
{\cal A}(\frac{u\bar{u}+d\bar{d}}{\sqrt{2}}\rightarrow \eta \eta ^{'}) \; : \;
{\cal A}(s\bar{s}\rightarrow K\bar{K}) \; : \;  {\cal A}(s\bar{s}\rightarrow 
\eta\eta ) \; : \;  {\cal A}(s\bar{s}\rightarrow \eta \eta ^{'}) \nonumber  \\
 &  &
= 2\;\! : \;\! 1\;\! : \;\! 2\sin ^2\theta \;\! : \;\! 2\sin 
\theta \cos \theta \;\! : \;\! \sqrt{2} \;\! : \;\! 2\sqrt{2}\cos ^2
\theta \;\! : \;\! -2\sqrt{2}\sin \theta \cos \theta .
\eeqna

When we take $\frac{f_{n\bar{n}G}}{m_{\tilde{G}}^2 - m_{n\bar{n}}^2}$ =
$\frac{f_{s\bar{s}G}}{m_{\tilde{G}}^2 - m_{s\bar{s}}^2}$ (explained below), we obtain
\beqna\plabel{nav}
\lefteqn{ \hspace{-4cm} {\cal M} (G\rightarrow \pi\pi) \; : \;  {\cal M} 
(G\rightarrow K\bar{K}) \; : \;  {\cal M} (G\rightarrow \eta \eta ) \; : \;  
{\cal M} (G\rightarrow \eta \eta ^{'}) =  \nonumber } \\ & &
1\; : \; 1\; : \; 1\; : \; 0.
\eeqna

This is the result one obtains when na\"{\i}vely coupling the quarks in the 
outgoing mesons to the vacuum \cite{amsler96}, often referred to as
``flavour democratic coupling''. It would also yield a 
horizontal line for our predicted amplitude in Fig. 1. 

The lattice results were obtained in the $SU(3)$ limit. To compare we shall
also adopt the $SU(3)$ limit in the remainder of this  
section. Hence we take the couplings and quark masses to be identical, i.e. 
$f_{n\bar{n}G} = f_{s\bar{s}G}\equiv f_{SU(3)G}$ and $m_{n\bar{n}}=m_{
s\bar{s}}\equiv m_{SU(3)}$.

Fig. 1 shows our results. 

The solid line represents our basic prediction. We use $\beta = 0.4$ GeV found
to enable a fit of a large range of  meson decays \cite{kokoski87,biceps}.
Mock meson phase space is employed since this enables a prediction
of $\cal M$ for all $p$, as can be done in lattice QCD. Since we work in the limit of 
$SU(3)$ symmetry, we take the mock meson phase space parameters
to be those of say $s\bar{s}$ mesons, i.e. $\tilde{M_A} = 1.49$ GeV,  
$\tilde{M_B},\tilde{M_C} = 0.85$ GeV (see Table 5 of ref. \cite{kokoski87}). 
We take the pair creation constant to be $\gamma _0=0.39$ \cite{kokoski87}.

Since all parameters are constrained, except for $f_{SU(3)G} / 
(m^2_{SU(3)}-m_{\tilde{G}}^2)$, we regard our fit in Fig. 1 as a 
one--parameter fit. 
This parameter sets the overall decay strength of the glueball, and does 
{\it not} influence the relative strengths of the various decay modes. 

The lattice results\footnote{GF11 predicts $\frac{{\cal M}}{m_\rho} = 0.834^{ 
+ 0.603}_{ - 0.579}, 2.654^{+ 0.372}_{ - 0.402}, 3.099^{ + 0.364}_{ - 0.423}$ 
for scalar glueball decay to $\pi \pi, K\bar{K}$ and $\eta \eta ,$ 
respectively \protect\cite{weindecay,weinpriv}, where $m_\rho$ is the mass of 
the $\rho$ meson. The predictions are given as a function of pseudoscalar 
mass, which we translate to the momentum of the outgoing mesons $p$ using 
conservation of energy: $p^2 = m_G^2/4 - m_{PS}^2$, where $m_{PS}$ is the 
relevant pseudoscalar meson mass.} for the $\eta 
\eta,\; K\bar{K}$ and $\pi \pi$ decay modes are 
plotted from left to right as the data points. It is 
non--trivial that our results are consistent. The fact that the prediction in 
Fig. 1 is not a horizontal line, as one na\"{\i}vely expects, indicates that 
the detailed dynamics of the flux--tube and $^3P_0$ models combined with the 
hypothesis of glueball dominance captures the correct strong interaction 
dynamics. This success is not shared by other models of glueball decay based 
on perturbative QCD decay dynamics, where the na\"{\i}ve pattern of Eq. 
(\ref{nav}) arises \cite{cui98}. The other points and lines in Fig. 1 indicate
parameter variations and are discussed in the caption of the figure.

We fit 
\beq
\Big| \frac{f_{SU(3)G}}{m_{\tilde{G}}^2-m_{SU(3)}^2}\Big| =0.34\pm 0.04.
\eeq

\subsection{Scalar glueball decay to two mesons}
Having predicted the scalar glueball decay to pseudoscalar mesons, we are now 
in a position to make the first predictions in the literature of the decay of 
the scalar glueball to non--pseudoscalar mesons.  

Since we have fitted Eq. (3.6) using mock meson phase space, we
again use this convention and hence $\gamma_0 = 0.39$ \cite{kokoski87}. 
We again use $\beta=0.4$ GeV and do the calculation in the $SU(3)$ limit with 
the $s\bar{s}$ mock meson masses\footnote{Taking the parameters for $n\bar{n}$
mesons, i.e. $\tilde{M_A} = 1.25$ GeV, $\tilde{M_B},\tilde{M_C} = 0.85$ GeV 
\protect\cite{kokoski87} would give scalar meson widths to $\pi \pi ,K\bar{K}$ 
and $\eta \eta \sim 20\%$ larger. A fit to the lattice data then yields a
different value for $f_{SU(3)G}/(m_{SU(3)}^2-m_{\tilde{G}}^2).$ The 
predictions for the glueball widths to
$\pi \pi ,K\bar{K}$ and $\eta \eta $ are identical, and the dominant width in
Table 1, to $a_1\pi $, is $1\%$ different from the value quoted.} 
\cite{kokoski87}. The results are indicated in Table 1. 
The primitive glueball amplitudes should be understood to be correct up to a 
sign.
The analytical expressions used for the amplitudes can be found in Appendix A.

\begin{table}[t]
\begin{center} \plabel{dectab}
\begin{tabular}{|c|c||c|c|c|c|c||c|}
\hline %------------------------
Decay Mode & Wave & 1.4 GeV & 1.5 GeV & 1.6 GeV & 1.7 GeV & 1.8 GeV & Width 
(MeV) \\
\hline %\hline %------------------------
$ \pi \pi          $ & S & 1.31   & 0.96  & 0.55  & 0.093 & -0.41 & 6 (6)  \\
$ K\bar{K}         $ & S & 2.8    & 2.7   & 2.0   &  1.6  & 1.0   & 81 (82) \\
$ \eta \eta        $ & S & 3.2    & 2.9   & 2.4   &  2.0  & 1.4   & 27 (27) \\
$ a_1\pi           $ & P &$10.7 p$&$11.2 p$&$11.7 p$&$12.0 p$&$12.3 p$& 177  
(67) \\
$ \pi (1300)\pi     $ & S &        & 1.1   & 1.4  & 1.8  &  2.3 & 23 (7)  \\
$ \rho \rho         $ & S &        &       & 3.3   & 3.0   & 2.6   & 61 (46) \\
$ \omega \omega     $ & S &        &       & 3.5   & 3.1   & 2.7   & 16 (12) \\
$ K^{\ast }\bar{K}^{\ast } $ & S &       &       &       &       & 4.1   &   \\
%$ K_1(1270)K        $ & P &        &$    p$&$    p$&$    p$&$    p$& ( ) \\
$ (\pi \pi )_S(\pi \pi )_S$ & S &$<$ 9.6   &$<$ 10.2 &$<$ 10.9 &$<$ 11.5 &
$<$12.1 &$<$ 490 ($<$ 160) \\
\hline %------------------------------
\end{tabular}
\end{center}
\caption{ Amplitudes for the decay of a scalar glueball to two 
mesons in GeV. For P--wave decays the linear momentum dependence is explicitly 
separated (with $p$ in GeV). 
The amplitudes for a 1.6 GeV glueball should be regarded as our 
predictions for the primitive glueball. The other amplitudes are to be used 
for calculation of the decays of the physical states. The widths (including
all partial waves) are listed in the final column for a 1.6 GeV glueball,
assuming that all resonances are narrow. 
All calculations are for mock meson phase space, except the widths in 
brackets in the last column, which are for relativistic phase space with
$\gamma_0$ chosen to agree with lattice QCD predictions for glueball decay to 
$\pi \pi ,\; K\bar{K}$ and $\eta \eta $. When both S-- and D--wave 
amplitudes are possible the amplitude is the S--wave amplitude. The ratio 
(D--wave amplitude)/($p^2$ S--wave amplitude) is -4.2 and -4.1 for decays of a 
1.6 GeV glueball to $\rho \rho $ and $\omega \omega ,$ respectively, and 
-4.0 for decay of a 1.8 GeV glueball to $K^{\ast }\bar{K}^{\ast }$.
$(\pi \pi )_S$ stands for a (hypothetical) narrow $(u\bar{u}+d\bar{d})/
\protect\sqrt{2}$ resonance $f_0(600)$ which decays dominantly to $\pi\pi$ 
\protect\cite{pdg}. It may be related to the low mass tail of the $f_0(1370) / f_0(400-1200)$.
Due to the large width and uncertain mass of the physical $(u\bar{u}+d\bar{d})
/\protect\sqrt{2}$ scalar resonance, the predictions should be viewed as being
anywhere between zero and the upper limit quoted.} 
\end{table}

We see from Table 1 that the total width of the 1.6 GeV scalar glueball is 
$250 - 390$ MeV excluding $(\pi \pi )_S(\pi \pi )_S$ decays. There is also
substantial phase space dependence for the glueball decay amplitudes to
$\pi \pi ,\; K\bar{K},\; \eta\eta$ and $\pi (1300)\pi $. 

\section{Glueball mixing}
\subsection{Glueball--quarkonium basis}
\subsubsection{Glueball--quarkonium linear mass matrix}
In ref. \cite{Wein}, Weingarten suggested the following $3\times 3$ linear 
mass matrix, which stems from the Hamiltonian formulation of QCD, to 
describe the mixing of a glueball and quarkonia:
\beq
\left(
\begin{array}{ccc}
m_G & z & \sqrt{2}z \\
z & m_{s\bar{s}} & 0 \\
\sqrt{2}z & 0 & m_{n\bar{n}}
\end{array}
\right) ,
\eeq
where $z$ stands for the annihilation amplitude of quarkonium into a
glueball which has dimensionality (mass) and represents a counterpart of
our $f$'s which have dimensionality (mass)$^2.$ In order to test our results 
by comparing with available lattice QCD data, we should establish a relation
between this linear mass matrix and our mass squared one. 

\subsubsection{Glueball--quarkonium quadratic mass matrix}
We first rewrite Weingarten's matrix for the squares of the glueball and 
quarkonia masses and show its equivalence to glueball dominance in the 
$2\times 2$ subspace spanned by quarkonia. We then establish a relation 
between Weingarten's linear and our quadratic mass matrices.

So, consider
\beq
\left(
\begin{array}{ccc}
m_G^2 & f & \sqrt{2}f \\
f & m_{s\bar{s}}^2 & 0 \\
\sqrt{2}f & 0 & m_{n\bar{n}}^2
\end{array}
\right) ,
\eeq
where the vanishing off--diagonal elements indicate that there is no direct
quarkonium--quarkonium mixing, i.e., glueball dominance.

{\bf Proposition 1.} The mass matrix (4.2) is equivalent (gives the same 
physical quarkonia masses) to glueball dominance 
in the $2\times 2$ subspace spanned  by quarkonia (with 
$f_{n\bar{n}G}=f_{s\bar{s}G}$).

{\bf Proof.} 
First, we rewrite the mass matrix (2.5) (with $f_{n\bar{n}G}=f_{s\bar{s}G}$) 
in the following form:
\beq
\left(
\begin{array}{cc}
m^2_{s\bar{s}}-r^2A & \sqrt{2}\;\!rA \\
\sqrt{2}\;\!rA & m^2_{n\bar{n}}-2A
\end{array}
\right) ,\;\;\;A\equiv 
\frac{f^2_{n\bar{n}G}}{m^2_{\tilde{G}}-m^2_{n\bar{n}}},\;\;r
\equiv \sqrt{\frac{m^2_{\tilde{G}}-m^2_{n\bar{n}}}{
m^2_{\tilde{G}}-m^2_{s\bar{s}}}}.
\eeq
where $r$ is a complex number.
The masses of the two physical states are now determined from the equations
\bqry
m^2_{f_0}+m^2_{f_0^{'}} & = & m^2_{n\bar{n}}+m^2_{s\bar{s}}-A\left( 2+r^2
\right), \\
m^2_{f_0}m^2_{f_0^{'}} & = & m^2_{n\bar{n}}m^2_{s\bar{s}}-A\left( 2m^2_{
s\bar{s}}+r^2m^2_{n\bar{n}}\right) .
\eqry

We take the equivalence of the matrices (2.5) and (4.2) to mean the equality 
of the corresponding eigenvalues $m^2_{f_0},\;m^2_{f_0^{'}}.$ 

The eigenvalues of (4.2) are determined from the following three equations:
\bqry
m_{\tilde{G}}^2+m_{f_0}^2+m_{f_0^{'}}^2 & = & m_G^2+m_{n\bar{n}}^2+m_{
s\bar{s}}^2, \\
m_{\tilde{G}}^2m_{f_0}^2+m_{\tilde{G}}^2m_{f_0^{'}}^2+m_{f_0}^2m_{f_0^{'}}^2 &
 = & m_G^2m_{n\bar{n}}^2+m_G^2m_{s\bar{s}}^2+m_{n\bar{n}}^2m_{s\bar{s}}^2-
3f^2, \\
m_{\tilde{G}}^2m_{f_0}^2m_{f_0^{'}}^2 & = & m_G^2m_{n\bar{n}}^2m_{s\bar{s}}^2-
f^2\left( 2m_{s\bar{s}}^2+m_{n\bar{n}}^2\right) ,
\eqry
%It then follows from (4.4),(4.6) that
%\beq
%m^2_{\tilde{G}}=m^2_G+A\left( 2+r^2\right) .
%\eeq
It then follows from Eqs. (4.4),(4.6) and Eqs. (4.5),(4.8) that:
\beq
f^2=\left( \frac{2m^2_{s\bar{s}}+r^2m^2_{n\bar{n}}}{2m^2_{s\bar{s}}+m^2_{
n\bar{n}}}\;\!m^2_{\tilde{G}}-\frac{(2+r^2)m^2_{n\bar{n}}
m^2_{s\bar{s}}}{2m^2_{s\bar{s}}+m^2_{n\bar{n}}}\right) A,
\eeq
and from  Eqs. (4.4)-(4.7):
\beq
f^2=\left( \frac{2+r^2}{3}\;\!m^2_{\tilde{G}}-\frac{2m^2_{n\bar{n}}+
r^2m^2_{s\bar{s}}}{3}\right) A.
\eeq
Using the definition of $r$ in Eq. (4.3), the equivalence of 
Eqs. (4.9) and (4.10) follows by simple algebra. Also, using 
the definition of $A$ in Eq. (4.3), and inserting it in either Eqs.
(4.9) or (4.10), we obtain that $\pm f= f_{n\bar{n}G} = f_{s\bar{s}G}$.
\eproof

This important result means that glueball dominance is nothing else but an
effective representation of the glueball--quarkonia mixing in the $2\times 2$ 
subspace spanned by quarkonia. The relation is only possible because both 
formulations descibe the physics in terms of the same basis states.

It is natural to define $f\equiv f_{n\bar{n}G} = f_{s\bar{s}G}$ , noting that 
Eq. (4.2) assumes $SU(3)$ symmetric couplings. This definition is 
consistent with what is obtained when the equivalence of the $2\times 2$ and 
$3\times 3$ formalisms is demanded in Proposition 1.

The mass matrix (4.2) possesses, however, more generality than the na\"{\i}ve 
glueball dominance picture in the $2\times 2$ quarkonia subspace.
This is because the former, in contrast to the latter, allows one to obtain 
the valence glue content of the physical quarkonia, and the valence content of
the physical glueball.

%If,
%e.g., $m_{n\bar{n}}<m_{s\bar{s}}\stackrel{<}{\sim }m_G,$
%the condition of Proposition 1 is only satisfied for $n\bar{n}$ but not for 
%$s\bar{s}:$ $f^2\gg (m_G^2-m_{s\bar{s}}^2)^2,$ 
%the masses of the physical states which diagonalize the mass matrix (4.2) are
%\bqry
%m_{\tilde{G}}^2 & \simeq  & \frac{m_G^2+m_{s\bar{s}}^2}{2}+|f|+ 
%\frac{2f^2}{m_G^2-m_{n\bar{n}}^2}, \\
%m_{f_0^{'}}^2 & \simeq  & \frac{m_G^2+m_{s\bar{s}}^2}{2}-|f|, \\
%m_{f_0}^2 & \simeq  & m_{n\bar{n}}^2-\frac{2f^2}{m_G^2-m_{n\bar{n}}^2}.
%\eqry
%which are approximations of the general expressions in this case.
%Note that because of the typical mass splitting $m_{s\bar{s}}-m_{n\bar{n}}=
%250\pm 50$ MeV \cite{pdg,BG}, the case $m_{n\bar{n}}\simeq m_{s\bar{s}}\simeq 
%m_G$ cannot be realized, and therefore is not considered here. 

{\bf Proposition 2.} The linear, (4.1), and quadratic, (4.2), formulations 
for the scalar mesons are equivalent provided that (i) $z^2\ll m^2_G,m^2_{
n\bar{n}},m^2_{s\bar{s}},$ (ii) $m_{s\bar{s}}-m_{n\bar{n}}\ll m_{n\bar{n}},
m_{s\bar{s}}$ or $m_{n\bar{n}},m_{s\bar{s}}\ll m_G.$

{\bf Proof.} We take the equivalence of the linear, (4.1), and quadratic, 
(4.2), formulations to mean (i) the equality of the eigenvalues of the matrix 
(4.2) to the eigenvalues squared of the matrix (4.1), (ii) the equality of the 
eigenvectors of both matrices, for the same values of the input parameters 
$m_G,m_{n\bar{n}},m_{s\bar{s}}.$ Denote the matrices (4.1),(4.2) by $M_{lin}$ 
and $M_{qaud},$ respectively, the corresponding diagonalized matrices by 
$\Lambda _{lin}$ and $\Lambda _{qaud},$ and the matrix that diagonalizes the 
$M$'s by $S$ (it is the same for $M_{lin}$ and $M_{qaud}$ because both have 
by construction the same eigenvectors).

Since $\Lambda _{lin}={\rm diag}\;(m_{\tilde{G}},m_{f_0^{'}},m_{f_0}),$ 
$$\Lambda ^2_{lin}={\rm diag}\;(m_{\tilde{G}}^2,m_{f_0^{'}}^2,m_{f_0}^2)=
\Lambda _{qaud}.$$
It follows from this relation and
$$\Lambda _{lin}^2=SM_{lin}S^{-1}\cdot SM_{lin}S^{-1}=SM^2_{lin}S^{-1},\;\;\;
\Lambda _{qaud}=SM_{qaud}S^{-1}$$ that
\beq
M_{qaud}=M^2_{lin}.
\eeq
Thus, the linear and quadratic formulations are equivalent provided that Eq. 
(4.11) is valid. 

Since the square of the mass matrix (4.1) is
\beq
\left(
\begin{array}{ccc}
m_G^2+3z^2 & z(m_G+m_{s\bar{s}}) & \sqrt{2}z(m_G+m_{n\bar{n}}) \\
z(m_G+m_{s\bar{s}}) & m_{s\bar{s}}^2+z^2 & \sqrt{2}z^2 \\
\sqrt{2}z(m_G+m_{n\bar{n}}) & \sqrt{2}z^2 & m_{n\bar{n}}^2+2z^2
\end{array}
\right) ,
\eeq
it is clear that Eq. (4.11) follows if both conditions (i) and (ii) of 
Proposition 2 are satisfied.
\eproof 

The equivalence of (4.2) and (4.12) also implies the following relation 
between $z$ and $f\!:$ $f=z(m_G+m_{q\bar{q}}).$ In the scalar sector where 
the glueball and quarkonia have comparable masses, it reduces to
\beq
f\simeq 2zm_G.
\eeq
For any other $J^{PC}$ multiplet, $m_G\gg m_{q\bar{q}},$ and, respectively, 
$f\simeq zm_G.$ 

An explicit numerical example on the equivalence of linear and quadratic
mass matrix formulations is given in Appendix B.

\subsection{Scalar meson spectroscopy}
As the relation between Weingarten's linear mass matrix and our quadratic mass 
matrix is established in the previous subsection, we are 
ready to consider scalar meson spectroscopy implied by glueball dominance,
and compare our results with the lattice QCD simulations of refs. 
\cite{weinmix,Wein}.

We shall first show that within the glueball dominance hypothesis, the $f_0(1500)$
cannot be the heaviest isoscalar scalar meson arising from ground state 
$n\bar{n},s\bar{s}$ and glueball mixing if $m_G>1.5$ GeV.\footnote{This 
conclusion may be modified by the inclusion of effects beyond glueball 
dominance, e.g., coupled channels \cite{pennington}, which may be
especially relevant for scalar states \cite{geigerloop}. Ref. \cite{pennington}
finds that the masses of the states are always lower than those of the 
primitive states. We believe this to be an artifact of the inclusion of only 
low--lying channels, which could lead to misleading results \cite{geigerloop}.}

The argument is as follows. There are three possibilities: %\footnote{The arguments in the 
%proof is also valid when using the exact relations (4.6)-(4.8).}
(i) $m_{s\bar{s}}<m_G,$ (ii) $m_{s\bar{s}}>m_G,$ (iii) $m_{s\bar{s}}=m_G.$
The main property of the $3\times 3$ mass matrices (4.1),(4.2) (we do not 
prove it here) is that upon mixing the higher mass primitive state becomes 
more massive, while the lower mass primitive state becomes less massive (i.e.,
the mass splitting between the higher and lower mass primitive states 
increases as a result of the mixing). Therefore, in the case (i) 
$m_{\tilde{G}}>m_G>1.5$ GeV; in the case (ii) $m_{f_0^{'}}>m_{s\bar{s}}>m_G>
1.5$ GeV. Finally, in the case (iii) it can be shown that the physical
$s\bar{s}$ and glueball states have masses $\simeq \sqrt{m_G^2\pm f},$ and 
therefore, one of them is always higher than $m_G > 1.5$ GeV. 

Hence, the $f_0(1500)$ is not the heaviest isoscalar scalar meson arising from
ground state $n\bar{n},s\bar{s}$ and glueball mixing. If both the existence of 
$f_0(980)$ and $f_0(1370) / f_0(400-1200)$ are confirmed by experiment, there has to be an 
extra degree of freedom to account for the existence of these states. 

\subsubsection{The glueball--quarkonium coupling $f$}
In ref. \cite{weinmix}, Lee and Weingarten estimate the mixing parameter $z$ 
to be\footnote{In a most recent paper \cite{LW1}, Lee and Weingarten introduce 
$SU(3)$ breaking effects, in terms of different values of $z$ for $G$--$n\bar{
n},$ $G$--$s\bar{s}$ mixing, $z_{Gn\bar{n}}/z_{Gs\bar{s}}= 1.198\pm 
0.072,$ and estimate $|z_{Gs\bar{s}}|=43\pm 31$ MeV, which implies, with the 
above ratio, $|z_{Gn\bar{n}}|=54\pm 40$ MeV, similar to (4.14).}
\beq
|z|=56\pm 37\;{\rm MeV.}
\eeq
Eq. (4.14) implies, via (4.13) with $m_G=1.65\pm 0.06$ GeV \cite{LW1},
\beq
|f|\simeq 0.19\pm 0.13\;{\rm GeV}^2.
\eeq
This implies that in the flavor $SU(3)$ limit where we take $m_{SU(3)} = m_{
n\bar{n}}=m_{s\bar{s}}=m_{s\bar{n}}=m_{K_0^\ast }=1.43$ GeV \cite{pdg} (and 
the same $m_G=1.65\pm 0.06$ GeV), 
\beq
\Big| \frac{f}{m_G^2-m_{SU(3)}^2}\Big| =0.28^{+0.38}_{-0.21},
\eeq
consistent with the value needed from glueball decay (Eq. (3.6)).

%On the other hand, the value of $f$ can be extracted by solving the system of 
%equations (4.16)-(4.18) below for $f,$ $m_{f_0}$ and $m_{f_0^{'}}$ fixing the 
%remaining third physical mass at 1.7 GeV and choosing the input parameters 
%$m_G,m_{n\bar{n}},m_{s\bar{s}}$ in agreement with (4.24). In doing so, we vary
%$m_{s\bar{s}}$ in the range $1.25-1.65$ GeV, and find $|f|$ varying from 
%$0.43\pm 0.01$ GeV$^2$ for $m_{s\bar{s}}=1.25$ GeV to $0.20\pm 0.00$ for 
%$m_{s\bar{s}}=1.65$ GeV. We also find that for $m_{s\bar{s}}=1.55$ MeV, 
%$|f|=0.30\pm 0.01$ GeV$^2$ and one of $f_0$'s has mass $1504\pm 4$ MeV, 
%consistent with $f_0(1500)$. The ratio $|f/(m_{\tilde{G}}^2-m_{n\bar{n}}^2)|$
%varies slowly from 
%$0.27\pm 0.01$ for $m_{s\bar{s}}=1.25$ GeV to $0.35\pm 0.08$ for 
%$m_{s\bar{s}}=1.65$ GeV, consistent with (4.22), while the ratio 
%$|f/(m_G^2-m_{s\bar{s}}^2)|$ is very sensitive to the value of $m_{s\bar{s}}$ 
%(e.g., it is infinite for $m_{s\bar{s}}=m_G=1.6$ GeV). 

We take from Eq. (3.6), defining $m_{q\bar{q}}=(m_{n\bar{n}}+m_{s\bar{s}})/2$,
\beq
\Big| \frac{f}{m_{\tilde{G}}^2-m_{q\bar{q}}^2}\Big| =0.34\pm 0.04.
\eeq
%which is the central value of (4.xx) consistent with the value needed from 
%glueball decay, as a relation between the scalar 
%glueball--quarkonium quadratic mass matrix, Eq. (4.2), and lattice QCD. 
%This relation will be used later on.

\subsubsection{Two simulations for glueball--quarkonium mixing}
We now wish to consider two simulations for the glueball--quarkonia mixing 
based on the quadratic mass matrix (4.2), and extract the masses of the 
primitive quarkonia and the physical states, with the help of Eqs. (4.15) and 
(4.17).

We use Eqs. (4.6)-(4.8) for the masses of the physical states $m_{\tilde{G}},$
$m_{f_0},$ $m_{f_0^{'}}.$

For the first simulation, we also employ Eq. (4.17) and
\beq
m_G=1.6\;{\rm GeV,}\;\;\;m_{s\bar{s}}-m_{n\bar{n}}=250\pm 50\;{\rm MeV,}
\eeq
the latter being a typical mass splitting\footnote{Note that we could use the 
mass squared splitting for quarkonia, $m_{s\bar{s}}^2-m_{n\bar{n}}^2,$ in 
place of the linear one in Eq. (4.18). With, e.g., the value for this mass 
squared splitting $0.65\pm 0.01$ GeV$^2,$ since for the remaining three 
$P$--wave nonets (in GeV$^2)$ $2(m_{K_2^{\ast }}^2-m_{a_2}^2)\simeq 0.64,$ 
$2(m_{K_{1A}}^2-m_{a_1}^2)\simeq 0.66,$ $2(m_{K_{1B}}^2-m_{b_1}^2)\simeq 0.65$ 
\cite{BG}, the solution is $m_{\tilde{G}}=1694\pm 27\;{\rm MeV,}$ $m_{f_0}=
1242\pm 21\;{\rm MeV,}$ $m_{n\bar{n}}=1314\pm 12\;{\rm MeV,}$ $m_{s\bar{s}}=
1542\pm 8\;{\rm MeV,}$ $|f|=0.285\pm 0.06\;{\rm GeV}^2,$ and $\sqrt{(m_{
n\bar{n}}^2+m_{s\bar{s}}^2)/2}=1432\pm 10$ MeV, consistent with the case of 
the linear mass splitting (middle column, Table 2).} between $s\bar{s}$ and 
$n\bar{n}$ states for different meson multiplets \cite{pdg,BG}, and solve the 
system of 5 equations (4.6)-(4.8),(4.17),(4.18) for $m_{\tilde{G}},$ $m_{f_
0},$ $m_{n\bar{n}},$ $m_{s\bar{s}}$ and $f,$ by fixing $m_{f_0^{'}}=1.5$ GeV. 

The reason for the latter requirement is that $f_0(1500)$ is established in 
more decay channels that any other scalar meson, and we should therefore 
construct our simulation of scalar meson spectroscopy with the constraint 
that one of the masses of the physical states is $1503\pm 11$ MeV \cite{pdg}.

For the second simulation, we employ Eq. (4.15) in place of (4.17).
The solution to Eqs. (4.6)-(4.8),(4.17),(4.18) for the first 
case, and Eqs. (4.6)-(4.8),(4.15),(4.18) for the second case are presented in 
Table 2.\footnote{Note that, although we do find a solution with $m_{s\bar{s}}
>m_G$ for both simulations, the values for $m_{n\bar{n}}$ and $m_{s\bar{s}}$ 
obtained are too high to be accommodated by any of the existing quark models; 
typically $m_{n\bar{n}}\sim 1.6$ GeV, $m_{s\bar{s}}\sim 1.8$ GeV. 
Also, $f_0(1500)$ is the lightest of the three scalars. Our results 
are therefore in agreement with the conclusion of ref. \cite{weinmix} that the
situation where the primitive $s\bar{s}$ state has a higher mass than the 
primitive glueball is incompatible with lattice QCD.} \\ 

\begin{center}
%{\footnotesize
\begin{tabular}{|c||c||c|} \hline
  & Eqs. (4.6)-(4.8),(4.17),(4.18) & Eqs. (4.6)-(4.8),(4.15),(4.18)  \\ \hline
 $m_{\tilde{G}},$ MeV & $1703\pm 40$ &  $1649^{+63}_{-41}$  \\ \hline
  $m_{f_0},$ MeV  & $1218\pm 70$ &  $1248^{+52}_{-72}$  \\ \hline
  $m_{s\bar{s}},$ MeV & $1546\pm 17$ &   $1527\pm 23$       \\ \hline
  $m_{n\bar{n}},$ MeV & $1296\pm 33$ &   $1277\pm 27$       \\ \hline
     $|f|,$ GeV$^2$     & $0.305\pm 0.09$ & $0.19\pm 0.13$   \\ \hline   
\end{tabular}
%}
\end{center}
Table 2: Solution to Eqs. (4.6)-(4.8),(4.17),(4.18), and Eqs. 
(4.6)-(4.8),(4.15),(4.18).
%\hspace*{0.64in}
\\

We note that the value of $f$ obtained in the first case is consistent with 
the value extracted from lattice QCD (Eq. (4.15)). It is also in agreement 
with values  extracted phenomenologically for different $J^{PC}$ meson nonets 
in ref. \cite{BBG} which all are in the interval $0.27-0.32$ GeV$^2.$ We 
however disagree with lattice QCD that the primitive $s\bar{s}$ is at least 
200 MeV below the primitive glueball \cite{weinmix,LW1} but only $\simeq 70\pm
30$ MeV, as seen from our solutions for $m_{s\bar{s}}$ in the two cases 
considered.
%Translating this $f$ into $z$ with the help of 
%(4.19) gives $|z|=93\pm 13$ MeV, which is consistent with (4.20). 
We also note that $\sqrt{(m_{n\bar{n}}^2+m_{s\bar{s}}^2)/2}$ which is $1426\pm
24$ and $1407\pm 25$ MeV in the two cases, respectively, is consistent with 
$m_{K_0^\ast }=1429\pm 6$ MeV \cite{pdg}.

With $f>0,$ the valence content of the three physical states obtained in the 
first simulation, is
\bqry
|1703\rangle \!\!\!& = & \;\!(0.821\!\pm \!0.02)|G\rangle \!+\!(0.493\!\pm \!
0.02)|s\bar{s}\rangle \!+\!(0.287\!\pm \!0.05)|\frac{u\bar{u}+d\bar{d}}{
\sqrt{2}}\rangle , \\
|1500\rangle \!\!\!\!& = & \!\!\!\!-(0.410\!\pm \!0.04)|G\rangle \!+\!(0.860\!
\pm \!0.02)|s\bar{s}\rangle \!-\!(0.305\!\pm \!0.08)|\frac{u\bar{u}+d\bar{d}}{
\sqrt{2}}\rangle , \\
|1218\rangle \!\!\!\!& = & \!\!\!\!-(0.397\!\pm \!0.08)|G\rangle \!+\!(0.133\!
\pm \!0.05)|s\bar{s}\rangle \!+\!(0.908\!\pm \!0.05)|\frac{u\bar{u}+d\bar{d}}{
\sqrt{2}}\rangle ,
\eqry
and shows that the physical glueball contains $\sim 70$\% glue and $\sim 30$\% 
$q\bar{q},$ while each of the physical quarkonia contains $\sim 15$\% glue and
$\sim 85$\% $q\bar{q}.$  The overall signs for the states have no physical 
significance.

Although the masses of the physical states do not depend on the sign of $f,$ 
the valence content of the physical states does. Namely, we find that under the
inversion of the sign of $f$ both the quark content of the physical glueball 
and the glue content of the physical quarkonia change their sign. However, 
it is not difficult to see that the $^3P_0$ model decay width of the physical 
states remains invariant under the inversion of the sign of $f.$ Hence, for 
the study of masses and decays of scalar mesons, one need to consider the case
$f>0$ only. 

Notwithstanding the similarity of the results
obtained, there is a principal difference between our approach and that of 
refs. \cite{weinmix,LW1,Wein}. Lee and Weingarten choose the input parameters 
$m_G,$ $m_{s\bar{s}},m_{n\bar{n}}$ and $z$ 
%(with $m_{n\bar{n}}$ fixed in agreement with the mass of the $a_0(1450))$ 
to obtain the three physical masses. Although the input parameters 
$m_G$ and $z$ obtained are consistent with their lattice QCD calculations, the
mass splitting $m_{s\bar{s}}-m_{n\bar{n}}\sim 50$ MeV, is atypical of that 
provided by constituent quark models in Eq. (4.18). Moreover, the 
Gell--Mann--Okubo type relation $m^2_{n\bar{n}}+m^2_{s\bar{s}}=2m^2_{s\bar{n}}=
2m_{K_0^{\ast }}^2$ \cite{BG} does not hold in their approach. In contrast, 
our approach does not assume $m_{u\bar{u}}$ and $m_{s\bar{s}}$ individually,
but only their difference. To compensate, we need one more input parameter:
we require that one of the physical masses be in agreement with the  
well established scalar state $f_0(1500)$.

\section{Phenomenology}

For some time there has been a controversy over the spin $J$ assignment of 
$f_J(1710)$, and hence the existence of a scalar state at this mass 
\cite{amsler96}. This matter is not yet fully resolved. Our results in Section
4.2 suggest that $f_0(1500)$ cannot be the heaviest state arising from 
glueball--quarkonium mixing. Hence we highlight recent evidence for a $J=0$ 
component in $f_J(1710)$. BES seperated both a $J=0$ and a $J=2$ component, 
with the scalar state having mass $1780$ MeV and a width of $85\pm 25$ MeV 
\cite{bes}. There are also claims of a $J=0$ signal at $1750\pm 15$ MeV with 
width $160\pm 40$ MeV \cite{mark3}, and with mass $1704^{+16}_{-23}$ MeV with 
width $124^{+52}_{-44}$ MeV \cite{dunwoodie}. $m_{\tilde{G}}$ obtained in 
Table 2 is consistent with these experimental masses.

%The mass of the lowest scalar meson in Table 2, $m_{f_0^{'}}$, is in agreement
%with the experimental value of $1.2-1.5$ GeV \cite{pdg} for the $f_0(1370) / f_0(400-1200)$,  
%although we cannot rule out its interpretation as the $f_0(980)$ 
%ained turn
%out in agreement with data on the scalar states at 1.7 and 1.3 GeV (the 
%inclusion of the coupling to the open channels in our analysis might shift the
%mass of the $f_0^{'}$ down from $\sim 1250$ MeV that we find, so that this 
%state could be consistent with Pennington's $f_0[\varepsilon (1000)]$ 
%\cite{pennington} or experimental $f_0(980)).$

The $a_1\pi$ decay of the primitive glueball is expected to be larger than
any pseudoscalar decay mode. There is some evidence for the production of
$f_0(1500)$ via $a_1^+$ exchange in the reaction $\pi^- p\rightarrow 
\pi^+ \pi^- n$, i.e. an $a_1\pi$ coupling of the $f_0(1500)$, in 
CERN--Cracow--Munich data with a polarized target \cite{lesniak}. 
We urge experimenters to allow for the $a_1\pi$ decay mode in partial wave 
analyses. This applies to Crystal Barrel at CERN for analysis of $\pi^+ \pi^-
\pi^0 \pi^0$ \cite{ulrike} and $\pi^+ \pi^- \pi^+ \pi^-$ data,
and to Mark III $\pi^+ \pi^- \pi^+ \pi^-$ data \cite{mark3}.

The $(\pi\pi)_S(\pi\pi)_S$ decay of the glueball can be very substantial, 
depending sensitively on the width and mass of the intermediate $f_0$ coupling
to $(\pi\pi)_S$ (see Table 1).
There are indeed indications of substantial $(\pi\pi)_S(\pi\pi)_S$ decay modes
in $f_0(1500)$ and $f_0(1710)$ \cite{pdg,mark3,cbar4pi}.

The two--photon widths of the various states provide stringent consistency 
checks for our results.  In the flavour $SU(3)$ limit the $\gamma\gamma$ width 
for 
a state $\alpha_G |G\rangle \; +\; \alpha_{s\bar{s}} |s\bar{s}\rangle \; + \;
\alpha_{n\bar{n}} |\frac{u\bar{u}+d\bar{d}}{\sqrt{2}}\rangle$ is proportional 
to
\beq \label{photon}
\left(
%\pm 6 \alpha_G \; \Big|\frac{f_{SU(3)G}}{m^2_{SU(3)}-m_{\tilde{G}}^2}\Big| + 
\alpha_{s\bar{s}} + \frac{5}{\sqrt{2}} \alpha_{n\bar{n}} \right) ^2
\eeq
where we have taken the charges of the quarks into account, and normalized the 
expression to be unity when the state is built purely from $s\bar{s}$. 
%Here we 
%have also assumed from glueball dominance that the primitive glueball can 
%decay, with an {\it ad hoc} sign which cannot be obtained from the fit in Eq. (3.6), to two photons via an intermediate scalar 
%meson. Note that primitive glueball decay to two photons is at variance with the usual assumption 
%\cite{amsler96,closeglue} (motivated from perturbative QCD) that the primitive
%glueball has negligible two--photon coupling. The latter case can be obtained 
%from Eq. (\ref{photon}) by putting $f_{SU(3)G} = 0$.

%If we take physical states built from pure primitive glueball, $s\bar{s}$ and 
%$\frac{u\bar{u}+d\bar{d}}{\sqrt{2}}$ states, we obtain the two--photon width 
%ratio $4.2:1:\frac{25}{2}$ using Eqs. (3.6) and (\ref{photon}). 
For the states 
in Eqs. (4.19)-(4.21) we obtain the two--photon width ratio for the states 
$f_0(1370), f_0(1500), f_0(1710)$ to be 
%$6 : 1 : 10$ or $17 : 0.4 : 0.03$  
%with the glueball--quarkonium coupling in Eq. (3.6), and
$11 : 0.05 : 2$.
% with no primitive glueball decay to two photons.

If one takes model--dependent estimates of two--photon widths of $s\bar{s}$,
i.e. 0.16 -- 0.20 keV \cite{twophot}, one observes that the two--photon width 
of $f_0(1500)$ is consistent with the experimental bound of $< 0.17$ keV 
\cite{lafferty}. Since $f_0(1370)$ is dominated by light quarks, our estimate 
for the two--photon width of $f_0(1370)$ is $\frac{2}{25}\; 
%(6-17)
11$ times the 
two--photon width of $\frac{u\bar{u}+d\bar{d}}{\sqrt{2}}$
($3.25 - 6.46$ keV \cite{twophot}). This is consistent $5.4\pm 2.3$ keV 
\cite{pdg} from experiment. There are currently no experimental estimates for 
$\gamma \gamma $ width of the $J=0$ component of $f_J(1710)$.

From Eqs. (4.19)-(4.21) it is clear that the glueball content of $f_J(1710)$ 
($J=0$) is significantly higher than that of $f_0(1500)$ and $f_0(1370)$, 
which are similar. This can be tested by evaluating the states' coupling 
to two gluons in the model of ref. \cite{amsler96}:
$BR(f_0(1710)\rightarrow gg) \geq 0.5$ \cite{li} and $BR(f_0(1500)\rightarrow 
gg) = 0.3 - 0.5$ \cite{farrar} or $0.64\pm 0.11$ \cite{li}. Since the
expectation for a glueball is that $BR(G\rightarrow gg) \geq 0.5$ 
\cite{farrar}, both $f_0(1710)$ and $f_0(1500)$ are consistent with a 
sizable glueball component, and with $f_0(1710)$ having a larger glueball 
component. Moreover, ref. \cite{closeglue} concludes that $f_0(1370)$ may have
some glueball admixture, smaller than $f_0(1500)$ and $f_0(1710)$, but is 
dominantly quarkonium, partially in agreement with our results. Predictions 
here are complicated by the large width of $f_0(1370)$ \cite{farrar}.

\section{Summary}
In this paper we suggest a coherent view at the scalar glueball as having the
following properties:
 
(i) A (physical) intermediate state in scalar $q\bar{q}$ 
annihilation--creation transitions (called ``glueball dominance'').

(ii) A state decaying to two mesons via an intermediate scalar meson.

(iii) A primitive  state which mixes with the primitive $n\bar{n}$ 
$(n=u,d)$ and $s\bar{s}$ quarkonia to form three physical scalar mesons. 

Three main assumptions are employed in this work: glueball dominance, 
SU(3) symmetry and the assumption that only
ground state quarkonia are relevant to scalar glueball mixing and decay.

As can be seen in Figure 1, glueball dominance together with the calculation 
of energy dependent couplings in the $^3P_0$/flux--tube model can account 
for the counterintuitive primitive glueball 
couplings to $\pi \pi ,\; K\bar{K}$ and $\eta \eta $ 
found in lattice QCD. In Table 1 a total glueball width of greater than
$250-390$ MeV with a dominant $a_1\pi$ decay of $70-180$ MeV is predicted.
Decay to $(\pi\pi)_S(\pi\pi)_S$ may also be significant as was
observed experimentally for $f_0(1500)$ and $f_J(1710)$. 

The quadratic mass matrices in the $2\times 2$ quarkonium and $3 \times 3$ 
glueball--quarkonium formulations are equivalent. The 
linear and quadratic $3 \times 3$ glueball--quarkonium mass matrices are 
equivalent under the requirements 
that (i) $z^2\ll m_G^2,m_{s\bar{s}}^2,m_{u\bar{u}}^2$ and 
(ii) $m_{s\bar{s}}-m_{n\bar{n}}\ll m_{n\bar{n}},
m_{s\bar{s}}$ or $m_{n\bar{n}},m_{s\bar{s}}\ll m_G.$ The conditions mentioned 
are always fulfilled in this work, and an illustrative example of 
the equivalence can be found in Appendix B. 

The $f_0(1500)$ is not the heaviest state arizing from glueball--quarkonium
mixing, implying that if the existence of both $f_0(980)$ and $f_0(1370) / f_0(400-1200)$
is confirmed, allowance should be made for an additional degree of freedom.
The glueball--quarkonium coupling extracted from our glueball decay
analysis is consistent with estimates from lattice QCD.

The mass of the physical glueball is consistent with the experimental
$f_J(1710)$. Experimental two--photon and $J/\psi$ radiative decay data
are consistent with the valence content predicted for the physical states.

\subsection*{Acknowledgments}
We wish to thank T. Barnes, V.M. Belyaev, D.V. Bugg, W. Dunwoodie, T. Goldman,
N. Isgur, W. Lee and D. Weingarten for valuable discussions during the 
preparation of this work.

\appendix
\section*{Appendix A: Glueball decay couplings}
The glueball decay amplitudes (evaluated in Table 1) to various outgoing
 states 
are (for $\beta_A = \beta_B = \beta_C \equiv \beta$ and identical quark masses)

%\beqn
$${\cal M} = \frac{f_{SU(3)G}}{m_{SU(3)}^2-m_{\tilde{G}}^2}\;
\sqrt{\frac{8m_G^2\tilde{M}_B\tilde{M}_C}{\tilde{M}_A}}\;
\sqrt{\frac{2}{\beta}}\; \pi^{\frac{3}{4}}\;\gamma_0\; \exp \;\!\{- 
\frac{p^2}{12\beta^2} \} \; \varpi ,$$
%\nonumber
%\eeqn
where

\beqna
\lefteqn{\hspace{.7cm}  \varpi = \frac{2^{\frac{9}{2}}}{3^2} \left( 1-\frac{
2}{9} (\frac{p}{\beta })^2\right) \left\{ \barr{c} \!\!\!1 \hspace{2.65cm}
\mbox{ for } \pi \pi ,\; K\bar{K},\; \eta \eta  \\
\frac{1}{\sqrt{3}} \hspace{2.5cm}\mbox{ for } 
\rho\rho,\; \omega\omega,\; K^{\ast}{\bar{K}}^{\ast} \earr \right.
\hspace{1.0cm}\mbox{(S--wave)}     \nonumber }  \\ & &
\varpi = - \frac{2^7}{3^{\frac{9}{2}}} (\frac{p}{\beta})^2 
\hspace{4.8cm}\mbox{ for } \rho\rho,\; \omega\omega,\; K^{\ast}{\bar{K}}^{\ast}
\hspace{1.3cm}\mbox{(D--wave)} \nonumber \\ & &
\varpi =  \frac{2^{\frac{11}{2}}}{3^3} \frac{p}{\beta} 
\hspace{5.52cm}\mbox{ for } a_1\pi \hspace{3.35cm}\mbox{(P--wave)} \nonumber \\
 & &
\varpi = \frac{2^4}{3^{\frac{7}{2}}}\left( 1+\frac{19}{18}(\frac{p}{\beta})^2 -
\frac{1}{27} (\frac{p}{\beta})^4\right)  \hspace{1.4cm}\mbox{ for } \pi(1300)
\pi \hspace{2.4cm}\mbox{(S--wave)}  \nonumber \\ & &
\varpi = \frac{2^{\frac{11}{2}}5}{3^4} \left( 1+\frac{19}{180}(\frac{p}{
\beta})^2 - \frac{1}{270} (\frac{p}{\beta})^4\right) \hspace{0.65cm}
\mbox{ for }f_0(600)f_0(600) \hspace{1.46cm}\mbox{(S--wave)}  \nonumber 
\eeqna

\appendix
\section*{Appendix B: Numerical example of the equivalence of linear and 
quadratic mass matrix formulations}
An illustrative example of how Proposition 2 works is the case 
analyzed by Weingarten \cite{Wein} where for the input masses $m_{n\bar{n}}=
1450$ MeV, $m_{s\bar{s}}=1516$ MeV, $m_G=1635$ MeV and $z=77$ MeV, the masses 
of the three physical states are $m_{\tilde{G}}=1710$ MeV, $m_{f_0^{'}}=1.5$ 
GeV and $m_{f_0}=1390$ MeV. To translate this linear mass case into our mass 
squared one we use the relation $f = z (m_G+m_{q\bar{q}})$ near Eq. 
(4.16) which, with $m_{q\bar{q}} = (m_{n\bar{n}}+m_{s\bar{s}})/2,$ gives $f
\simeq 0.24$ GeV$^2.$ When this $f$ and the same input masses squared are used
in (4.2), we obtain the masses of the physical states, $m_{\tilde{G}}=1706$ 
MeV, $m_{f_0^{'}}=1.5$ GeV, $m_{f_0}=1383$ MeV, which are in excellent 
agreement with Weingarten's case, and the valence content of the physical 
states,
\bqryn
|1706\rangle & = & \;\;\;0.87|G\rangle +0.34|s\bar{s}\rangle +0.36|\frac{
u\bar{u}+d\bar{d}}{\sqrt{2}}\rangle , \\
|1500\rangle & = & -0.18|G\rangle +0.89|s\bar{s}\rangle -0.41|\frac{
u\bar{u}+d\bar{d}}{\sqrt{2}}\rangle , \\
|1383\rangle & = & -0.46|G\rangle +0.29|s\bar{s}\rangle +0.84|\frac{
u\bar{u}+d\bar{d}}{\sqrt{2}}\rangle ,
\eqryn
which is in excellent agreement with the corresponding valence content of ref. 
\cite{Wein}: 
\bqryn
|1710\rangle & = & \;\;\;0.87|G\rangle +0.34|s\bar{s}\rangle +0.36|\frac{
u\bar{u}+d\bar{d}}{\sqrt{2}}\rangle , \\
|1500\rangle & = & -0.19|G\rangle +0.90|s\bar{s}\rangle -0.40|\frac{
u\bar{u}+d\bar{d}}{\sqrt{2}}\rangle , \\
|1390\rangle & = & -0.46|G\rangle +0.28|s\bar{s}\rangle +0.84|\frac{
u\bar{u}+d\bar{d}}{\sqrt{2}}\rangle .
\eqryn

%We note that the example considered is also in agreement with
%the most recent calculation of ref. \cite{LW1} where small $SU(3)$ breaking
%effects, in terms of different values of $z$ for $G$--$n\bar{n},$ $G$--$s\bar{
%s}$ mixing, $z_{Gn\bar{n}}/z_{Gs\bar{s}}\approx 1.2,$ are taken into 
%account, and for the input masses $m_{n\bar{n}}=1470$ MeV, $m_{s\bar{s}}=1514$
%MeV, $m_G=1622$ MeV and $z_{Gs\bar{s}}=64$ MeV, the masses of the physical 
%states are $m_{\tilde{G}}=1697$ MeV, $m_{f_0}=1505$ MeV, $m_{f_0^{'}}=1404$ 
%MeV, and their valence content is in agreement with the above two cases.   

%We, of course, can arrange our 
%results to be in quantitative agreement with Lee and Weingarten: with $m_G=
%1646$ MeV, $m_{n\bar{n}}=1449$ MeV, $m_{s\bar{s}}=1514$ MeV and $|f|=0.208$ 
%GeV$^2$ (which corresponds to $|z|=63$ MeV and agrees with (4.23): $|f/(
%m_G^2-m_{n\bar{n}}^2)|=0.34),$ our physical masses are 1.4, 1.5 and 1.7 GeV, 
%and the valence content is in agreement with that of refs. \cite{Wein,LW1}. 

\hskip 1.9in    
\begin{center}
\vspace{2cm} 
\parbox{6in}{
\small Fig.~1:\quad The amplitude ${\cal M}$ (in GeV) plotted against the
square of the momentum in the outgoing
 state $p^2$ (in GeV$^2$). The sold line is 
our basic prediction and the data points are the lattice predictions, both of 
which are discussed in the text. We allow the following parameter variations. 
(a) The dashed line differs from the basic prediction in that we take inverse 
radii motivated from studies of masses and wave functions of mesons and 
glueballs, {\it not} decays. The $\pi$ and $K$ inverse radii were estimated as
0.54 GeV and 0.53 GeV respectively \cite{swanson92}, so we take $\beta _B=
\beta _C=0.54$ GeV. We also talk the glueball to have an r.m.s. radius 
$\sqrt{\langle r^2 \rangle}$ of $\sim \frac{1}{2}$ fm 
\protect\cite{bali93,gluesize}. Assuming that the scalar (P--wave) meson 
coupling to the glueball has the same size, one estimates for S.H.O. wave 
functions that $\beta_A \sqrt{\langle r^2 \rangle} = \sqrt{\frac{5}{2}}$, 
yielding $\beta_A = 0.6$ GeV. (b) The solid black dots differ from the basic 
prediction, in that we adopt the relativistic phase space convention 
\protect\cite{biceps,geiger94}. Here we make the replacements $\tilde{M}_A 
\rightarrow m_G$ and $\tilde{M}_B,\tilde{M}_C\rightarrow \sqrt{m_{PS}^2+p^2}$ 
in Eq. (\ref{coupling}), where $m_{PS}$ is the outgoing pseudoscalar mass. 
From left to right the points correspond to the $\eta\eta$, $K\bar{K}$ and 
$\pi\pi$ decay modes. The large points correspond to $\gamma_0 =0.4$ 
\protect\cite{biceps}, and the small points to  $\gamma_0 =0.53$ 
\protect\cite{geiger94}. } 
\epsfig{file=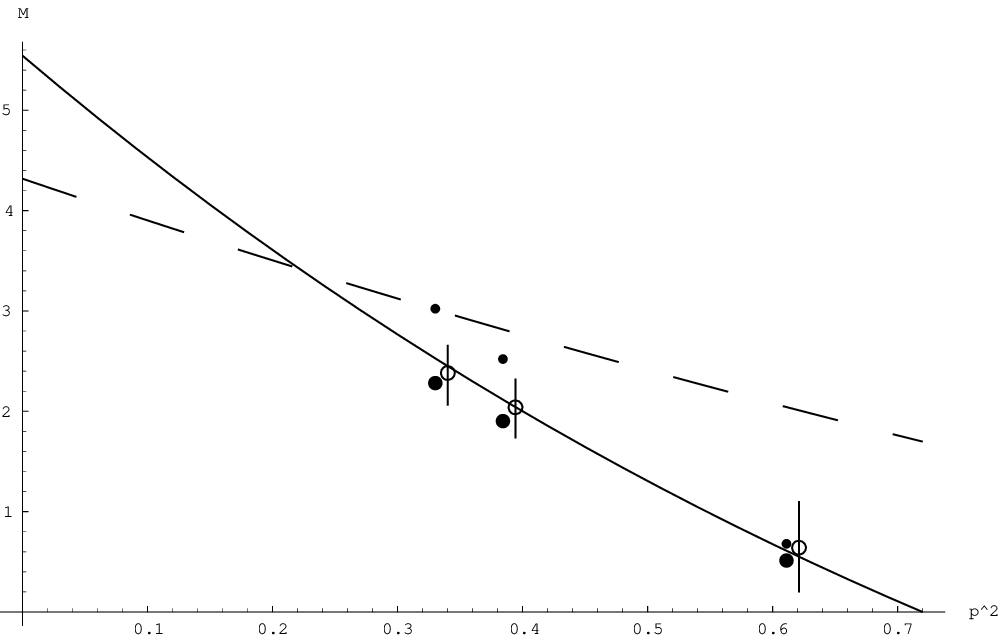,width=15cm,angle=-90}
\end{center}

\end{document}